\documentclass[aps,prl,a4paper,twocolumn,showpacs,floatfix]{revtex4}
\usepackage[dvips]{epsfig}
\begin{document}
\title{Ultracold collisions involving heteronuclear alkali metal dimers}
\author{Marko T. Cvita\v{s}}
\author{Pavel Sold\'{a}n}
\author{Jeremy M. Hutson}
\affiliation{Department of Chemistry, University of Durham, South Road,
Durham, DH1~3LE, England}
\author{Pascal Honvault}
\author{Jean-Michel Launay}
\affiliation{UMR 6627 du CNRS, Laboratoire de Physique des Atomes,
Lasers, Mol\'ecules et Surfaces, Universit\'e de Rennes, France}

\date{\today}

\begin{abstract}
We have carried out the first quantum dynamics calculations on
ultracold atom-diatom collisions in isotopic mixtures. The systems
studied are spin-polarized $^7$Li + $^6$Li$^7$Li, $^7$Li +
$^6$Li$_2$, $^6$Li + $^6$Li$^7$Li and $^6$Li + $^7$Li$_2$.
Reactive scattering can occur for the first two systems even when
the molecules are in their ground rovibrational states, but is
slower than vibrational relaxation in homonuclear systems.
Implications for sympathetic cooling of heteronuclear molecules
are discussed.
\end{abstract}
\pacs{03.75.Ss,33.80.Ps,34.20.Mq,34.50.Lf}

\maketitle

\font\smallfont=cmr7

There is at present great interest in the properties of cold and
ultracold heteronuclear molecules. There are several reasons for
this. First, there are new phases of matter predicted for
quantum-degenerate dipolar gases. Secondly, heteronuclear
molecules can be fermionic whereas neutral homonuclear molecules
cannot. Dipolar molecules have potential applications in quantum
computing and in the emerging field of ultracold chemistry.

Heteronuclear alkali metal dimers, which can be created from
ultracold atomic gas mixtures, are of particular interest.
Formation of such dimers by photoassociation has already been
detected in cold atomic traps
\cite{Big99,Zimm01,DeMille04a,Marcassa04,DeMille04b,Big04,Stwalley04}.
Very recently, magnetically tunable Feshbach resonances have been
observed experimentally in two such systems \cite{Kett04,Jin04}.
Efforts are under way to use these for production of ultracold
molecules. Cold bialkali dimers have also been formed on helium
droplets \cite{Wide04}.

The molecules (heteronuclear or homonuclear) produced so far from
ultracold atomic gases have been formed in high vibrational
states. These are prone to undergo inelastic collisions
(quenching) with the remaining atoms or other molecules; the
resulting recoil produces trap loss. In the case of homonuclear
molecules, this difficulty was overcome by making use of fermionic
isotopes ($^{6}$Li and $^{40}$K), and tuning the scattering length
to very large positive values. Pauli blocking then suppressed the
inelastic collisions significantly, and long-lived molecular
Bose-Einstein condensates were produced at the end of 2003
\cite{Grimm03a,Jin03a,Kett03a}.

Efforts are being made to produce alkali metal dimers in low-lying
vibrational states, either directly or by transferring the
population from high-lying states. However, we have shown
computationally that Pauli blocking does {\em not} suppress
inelastic collisions for low-lying vibrational states
\cite{Cvi05a,Que05a}. For such states the quenching rates dominate
the elastic rates below 1 mK even when fermionic isotopes are
used. To trap homonuclear alkali metal dimers with small bond
lengths, it will therefore be essential to produce them in their
ground rovibrational state.

In the case of heteronuclear molecules the situation is even more
complicated. Even if heteronuclear dimers are produced in their
ground rovibrational states, the molecules may not be stable
against collisions. For example, the spin-polarized reaction
\begin{equation}
^6\textrm{Li}^7\textrm{Li}(v=0,n=0) + {}^7\textrm{Li} \rightarrow
{}^6\textrm{Li} + {}^7\textrm{Li}_{2}(v=0,n=0) \label{react1}
\end{equation}
is exothermic by 1.822~K because of the difference between the
zero-point energies of the two dimers.

The focus of this Letter is the collision of lithium atoms with
lithium dimers in cold atomic gases containing both the $^6$Li and
$^7$Li isotopes. We have carried out the first full quantum
dynamics calculations for spin-polarized collisions involving
heteronuclear alkali metal dimers. We find that quenching via
process (\ref{react1}), is significantly slower than the
vibrational relaxation processes studied previously in homonuclear
collisions \cite{Cvi05a}. In the present case the elastic cross
section dominates the inelastic cross sections in a region between
10 $\mu$K and 10 mK.

Quantum scattering calculations require accurate potential energy
surfaces. We have previously shown that the interaction potential
for spin-polarized (quartet) Li + Li$_2$ is highly non-additive
\cite{Sol03}, with a well depth 4 times that of the sum of three
Li-Li triplet pair potentials. In the present work, we use a
global quartet Li$_3$ potential obtained from all-electron
coupled-cluster electronic structure calculations. This is the
same potential as was used in our work on homonuclear Li + Li$_2$
collisions \cite{Cvi05a} and will be described in detail elsewhere
\cite{Cvi05pot}.

The methods we use to carry out quantum dynamics calculations on
systems of this type have been described in our previous work on Na
+ Na$_2$ \cite{Sol02,Quem04}, and elsewhere in the context of
thermal reactive scattering \cite{hon04}, so a brief summary will
suffice here. The potential energy surface is barrierless, so that
it is essential to take reactive (atom exchange) collisions into
account. The positions of the nuclei are described in hyperspherical
democratic coordinates. The configuration space is divided into
inner and outer regions, and the boundary between them is placed at
a distance such that couplings due to the residual atom-diatom
interaction can be neglected outside the boundary. In the inner
region (hyperradius $\rho \le 45\ a_0$ in the present case), we
obtain the wavefunction for nuclear motion by propagating a set of
coupled equations in a diabatic-by-sector basis that is obtained by
diagonalizing a fixed-$\rho$ reference Hamiltonian in a basis set of
pseudohyperspherical harmonics. In the outer region, we use the
Arthurs-Dalgarno formalism \cite{Dalgarno}, which is based on Jacobi
coordinates, and compute by inwards integration regular and
irregular solutions of a radial Schr{\"o}dinger equation which
includes the isotropic ($R^{-6}$) part of the interaction. Matching
between wavefunctions in the inner and outer regions yields the
scattering S-matrix.

The present calculations deal with systems containing two nuclei
that are identical and a third that is different. We consider
spin-stretched states of all three atoms involved (states with
$|M_F|=F=F_{\rm max}$), for which the nuclear spin wavefunction is
symmetric with respect to any exchange of nuclei. However, the
electronic wavefunction for the quartet state is antisymmetric
with respect to exchange of nuclei because it depends
parametrically on nuclear positions. Thus the wavefunction for
nuclear motion must be antisymmetric with respect to exchange of
the two identical bosonic nuclei (for fermionic alkali metal
atoms) and symmetric with respect to exchange of the two identical
fermionic nuclei (for bosonic alkali metal atoms).

Asymptotically, the hyperspherical functions correlate with
atom-diatom functions; the diatom functions for Hund's case (b),
which is appropriate for Li$_2\ (^3\Sigma_u^+$), are labelled with
a vibrational quantum number $v$ and a mechanical rotational
quantum number $n$; $n$ couples with the diatomic electron spin
$s=1$ to give a resultant $j$, but for Li$_2$ the splittings
between states of the same $n$ but different $j$ are very small
and are neglected here. In the $^3\Sigma_u^+$ state, only even $n$
is allowed for $^7$Li$_2$ and only odd $n$ for $^6$Li$_2$. In the
outer region, we include rovibrational states with $v=0,1,...,7$
with all rotational levels up to $n_{\rm max}=32, 29, 26, 23, 20,
17, 13, 7$ for $^7$Li$^6$Li, with all even rotational levels up to
$n_{\rm max}=32, 30, 28, 24, 22, 18, 14 ,10$ for $^7$Li$_2$, and
with all odd rotational levels up to $n_{\rm max}=31, 27, 25, 21,
19, 15, 11, 7$ for $^6$Li$_2$. Because of the reduced symmetry,
the present calculations are substantially more expensive than our
previous ones on collisions involving 3 identical atoms, and we
have restricted them to total angular momentum and parity $J^\Pi =
0^+$ and $1^-$. The basis sets of pseudohyperspherical harmonics
involved between 3660 and 6488 functions.

There are four collision systems of interest to $^6$Li/$^7$Li
mixtures: \begin{eqnarray} ^6{\rm Li} &+& {}^6{\rm Li}^7{\rm Li} \label{en1}\\
^7{\rm Li} &+& {}^6{\rm Li}^7{\rm Li} \label{ex1}\\
^6{\rm Li} &+& {}^7{\rm Li}_2 \label{en2}\\
^7{\rm Li} &+& {}^6{\rm Li}_2. \label{ex2} \end{eqnarray}
%
We have calculated the low-energy elastic, inelastic and reactive
cross sections $\sigma(E)$ for all four of these collision
systems, considering all energetically allowed processes for a
variety of initial diatomic rovibrational states $(v,n)$. The
present paper will focus on collisions with the diatomic molecule
initially in its lowest allowed rovibrational state. Details of
the results for other initial states may be found in ref.\
\onlinecite{CviPhD}.

\begin{figure} [htbp]
\begin{center}
\rotatebox{270}{ \resizebox{7cm}{!}
{\includegraphics{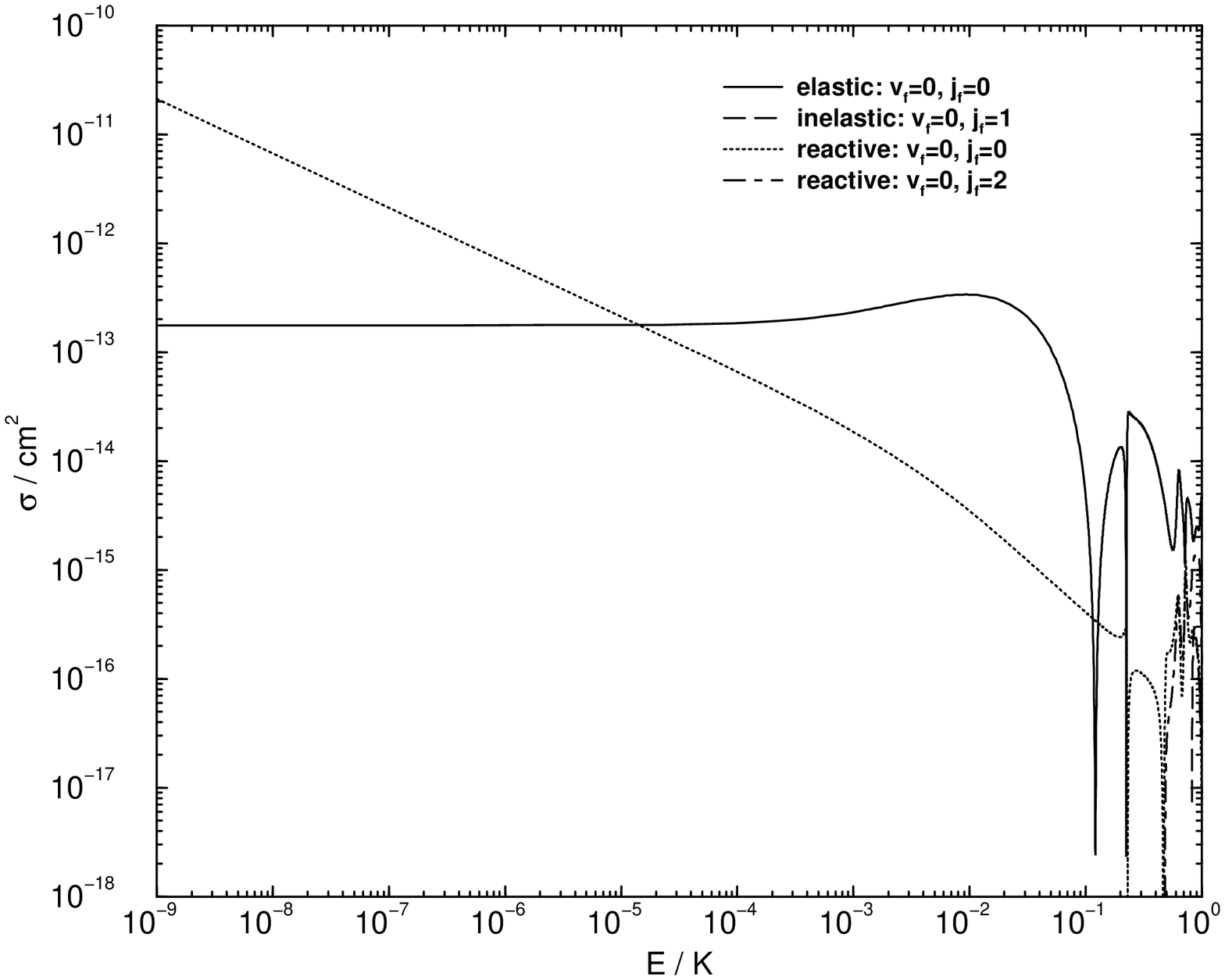}}} \caption{Elastic and reactive
s-wave cross sections for $^7$Li + $^6$Li$^7$Li$(v=0,n=0)$}
\label{xsec-7-67}
\end{center}
\end{figure}

At very low energies (up to about 1 mK), the cross sections are
dominated by collisions with orbital angular momentum $l=0$, for
which there are no centrifugal barriers. The s-wave cross sections
for $^7$Li + $^6$Li$^7$Li$(v=0,n=0)$ are shown as a function of
collision energy $E$ in Fig.\ \ref{xsec-7-67}. There is an
energetically allowed reactive pathway to form $^7$Li$_2(v=0,n=0)$
+ $^6$Li; at collision energies above 0.477~K the $v=0,n=2$ state
of $^7$Li$_2$ is also accessible.

Scattering calculations for Li + Li$_2$ are converged up to about
10$^{-4}$~K using just a single partial wave. At higher energies,
other partial waves start to contribute significantly. It may be
seen in Fig.\ \ref{xsec-7-67} that the reactive cross section for
$^7$Li + $^6$Li$^7$Li$(v=0,n=0)$ at 10$^{-4}$~K is $6.7 \times
10^{-14}$ cm$^2$, corresponding to a limiting low-temperature rate
coefficient $k_{\rm loss} = 4.1 \times 10^{-12}$ cm$^3$ s$^{-1}$.
This is about 2 orders of magnitude smaller than obtained
previously for $^6$Li + $^6$Li$_2(v>0)$ and $^7$Li +
$^7$Li$_2(v>0)$ \cite{Cvi05a}; the difference can be attributed to
the smaller kinetic energy release and single product
rovibrational channel in the present case (and is present only for
the initial state $v=0,n=0$). As a result, elastic scattering
dominates reactive scattering for $^7$Li + $^6$Li$^7$Li$(v=0,n=0)$
at collision energies between 10 $\mu$K and 10~mK. (The upper
limit is uncertain because the partial wave sum is not converged
at higher energies).

\begin{figure} [htbp]
\begin{center}
\rotatebox{270}{ \resizebox{7cm}{!}
{\includegraphics{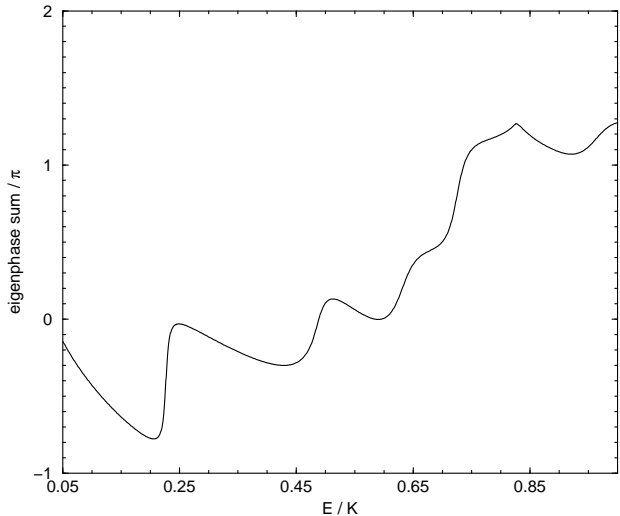}}} \caption{Eigenphase sum
for $^7$Li + $^6$Li$^7$Li scattering ($J^\Pi=0^+$).}
\label{epsum-7-67}
\end{center}
\end{figure}

The existence of a low-energy closed product channel increases the
density of low-energy Feshbach resonances. These are most
conveniently detected by plotting the S-matrix eigenphase sum
$\varsigma(E)$ \cite{Ash83} as a function of energy: at a
resonance, $\varsigma(E)$ increases sharply by $\pi$, superimposed
on a falling background. The eigenphase sum for $^7$Li +
$^6$Li$^7$Li is shown as a function of energy above the $v=0,n=0$
threshold in Fig.\ \ref{epsum-7-67}. There are clear isolated
resonances at 225 and 470~mK and a pair of overlapping resonances
at about 630 and 725~mK. The cusp at 825~mK corresponds to the
opening of the $(v=0,j=1)$ reactant channel. Small changes in the
potential energy surface might result in significant changes in
the positions of individual resonances, but the {\it density} of
resonances obtained here should be realistic. However, it should
be remembered that the present calculations neglect nuclear spin
and magnetic fields, which produce additional resonance structures
and ways to tune them.

For $^6$Li + $^6$Li$^7$Li$(v=0,n=0)$, the reactive channel to form
$^6$Li$_2$ is closed at energies below 0.825 K and only elastic
scattering is possible. The low-energy limit of the cross section
is $4.7 \times 10^{-12}$ cm$^2$.

\begin{figure} [htbp]
\begin{center}
\rotatebox{270}{ \resizebox{7cm}{!}
{\includegraphics{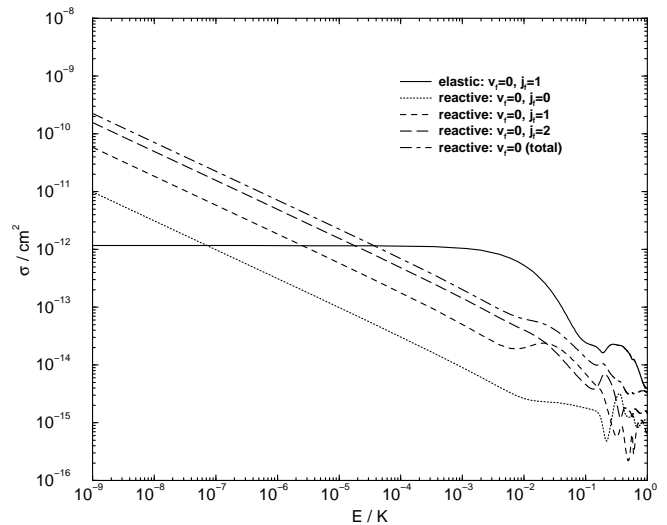}}} \caption{Elastic and reactive
cross sections for $^7$Li + $^6$Li$_2(v=0,n=1)$ with
$J^\Pi=1^-$.}\label{xsec-7-66}
\end{center}
\end{figure}

For $^6$Li$_2$, the lowest allowed rotational level is $n=1$. For
$^7$Li + $^6$Li$_2(v=0,n=1)$, the lowest partial wave to include
an $l=0$ incoming channel is therefore $J^\Pi=1^-$. The
corresponding cross sections are shown as a function of collision
energy $E$ in Fig.\ \ref{xsec-7-66}. The species can either
collide elastically or react exothermically to form $^6$Li$^7$Li +
$^6$Li with the diatom in its (0,0), (0,1) and (0,2) states. In
this case the reactive cross section at 10$^{-4}$ K is $7.1 \times
10^{-13}$ cm$^2$, which corresponds to a limiting low-temperature
rate coefficient $k_{\rm loss}=4.4 \times 10^{-11}$ cm$^3$
s$^{-1}$. This is considerably larger than for $^7$Li +
$^6$Li$^7$Li$(v=0,n=0)$, though the {\em partial} cross section to
form the $(v=0,n=0)$ product is comparable; the larger value for
the {\it total} reactive cross section reflects the larger kinetic
energy release and increased number of open product channels for
$^7$Li + $^6$Li$_2(v=0,n=1)$.

For $^6$Li + $^7$Li$_2(v=0,n=0)$ the reactive channel to form
$^6$Li$^7$Li is closed at energies below 1.822 K and only elastic
scattering is possible. The low-energy limit of the cross section
is $1.3 \times 10^{-12}$ cm$^2$.

The results obtained here have important implications for
experiments aimed at producing a quantum gas of $^6$Li$^7$Li in
its ground rovibronic state in an ultracold mixture of $^6$Li and
$^7$Li atoms. In order to stabilize the molecular cloud against
two-body trap losses induced by the reactive process
(\ref{react1}), the remaining atomic $^7$Li must be removed
quickly after ground-state molecule production, so that just the
two-species fermionic mixture of $^6$Li$^7$Li$(v=0,n=0)$ molecules
and $^6$Li atoms is left in the trap. One could separate the
$^6$Li atomic cloud from the molecular cloud as well, but it might
be advantageous to keep them both in the trap. It has been
observed that in a one-component dilute Fermi gas the s-wave
collisions are strongly suppressed because of Pauli blocking
\cite{Jin01}. This will make evaporative cooling of the molecular
cloud very inefficient. However, to achieve further cooling of the
$^{6}$Li$^{7}$Li$(v_{i}=0,n_{i}=0)$ molecules, a sympathetic
cooling scheme might be developed. Sympathetic cooling of
fermionic molecules can be approached from two directions:
interstate or interspecies sympathetic cooling.

Interstate sympathetic cooling is based on s-wave collisions
between particles of the same species but in two different states.
It was previously used, for example, to produce two overlapping
Bose-Einstein rubidium condensates \cite{Wieman97} and to create a
Fermi sea from potassium atoms \cite{Jin99}. Unfortunately, for
$^6$Li$^7$Li there are various inelastic processes resulting from
reactive molecule + molecule s-wave collisions that are likely to
produce significant trap losses. It will therefore be better to
keep the molecules in the same state, where Pauli blocking
suppresses such processes.

Interspecies sympathetic cooling makes use of s-wave collisions
between two different species. It has been used to produce a
Bose-Einstein condensate of atomic potassium \cite{Ing01} and to
create a two-species mixture of quantum-degenerate Bose and Fermi
gases \cite{Kett02}. In the present case we can distinguish two
interspecies sympathetic cooling schemes.

The first scheme uses the fact that ultracold collisions between
spin-stretched $^6$Li$^7$Li$(v=0,n=0)$ and $^6$Li can result only
in elastic scattering, because the corresponding inelastic and
reactive processes are energetically forbidden. Thus no
undesirable quenching-induced trap loss occurs. The elastic
collisions will then lead to sympathetic cooling between the
atomic and molecular clouds. This could be done in the trap where
the molecules were produced, or alternatively the $^6$Li$^7$Li
molecules could be transported into a reservoir of colder $^6$Li
atoms. This scheme is similar to interstate sympathetic cooling
for fermions in the sense that neither of the components can be
cooled evaporatively and the cooling is done only by interspecies
collisions.

If the efficiency of the first scheme is not high enough to
achieve quantum degeneracy in the molecular cloud, then an
alternative will need to be adopted. One possibility is to
transport the molecular cloud into a reservoir of bosonic atoms
such as $^{87}$Rb. Spin-polarized ultracold collisions between
$^6$Li$^7$Li$(v=0,n=0)$ and an alkali metal atom heavier than Li
can lead only to elastic scattering, because the reactive
quenching channels are not energetically accessible. The bosonic
component can then be cooled evaporatively. Note however that this
will work only if both species are in spin-stretched states, as
collisions involving other states could produce reaction to form
singlet states of molecules such as LiRb.

\acknowledgments

We are grateful to C. S. Adams and S. L. Cornish for valuable
discussions. PS and JMH are grateful to EPSRC for support under
research grant GR/R17522/01. MTC is grateful for sponsorship from
the University of Durham and Universities UK.

\end{document}